\documentclass[acmsmall]{acmart}

\AtBeginDocument{%
  }

\setcopyright{cc}
\setcctype{by}
\acmJournal{PACMHCI}
\acmYear{2025} \acmVolume{9} \acmNumber{7} \acmArticle{CSCW460} \acmMonth{11}
\acmDOI{10.1145/3757641}

\usepackage{graphicx}
\usepackage[normalem]{ulem}
\usepackage[utf8]{inputenx}
\usepackage{color,soulutf8}
\usepackage{multirow}

\usepackage{booktabs}
\usepackage{tabularx}
\usepackage{xcolor}
\usepackage{hyperref}

\begin{document}

\title[Organization Matters]{Organization Matters: A Qualitative Study of Organizational Dynamics in Red Teaming Practices for Generative AI
}

\author{Bixuan Ren}
\email{bren04@syr.edu}
\orcid{0009-0001-2706-3672}
\affiliation{%
  \institution{Syracuse University}
  \city{Syracuse}
  \state{New York}
  \country{USA}
}

\author{EunJeong Cheon}
\email{echeon@syr.edu}
\orcid{0000-0002-0515-6675}
\affiliation{%
  \institution{Syracuse University}
  \city{Syracuse}
  \state{New York}
  \country{USA}
}

\author{Jianghui Li}
\email{jli159@syr.edu}
\orcid{0009-0002-9800-6859}
\affiliation{%
 \institution{Syracuse University}
 \city{Syracuse}
 \state{New York}
 \country{USA}
 }

\renewcommand{\shortauthors}{Bixuan Ren, EunJeong Cheon, and Jianghui Li}

\begin{abstract}
  {The rapid integration of generative artificial intelligence (GenAI) across diverse fields underscores the critical need for red teaming efforts to proactively identify and mitigate associated risks. While previous research primarily addresses technical aspects, this paper highlights organizational factors that hinder the effectiveness of red teaming in real-world settings. Through qualitative analysis of 17 semi-structured interviews with red teamers from various organizations, we uncover challenges such as the marginalization of vulnerable red teamers, the invisibility of nuanced AI risks to vulnerable users until post-deployment, and the lack of user-centered red teaming approaches. These issues often arise from underlying organizational dynamics, including organizational resistance, organizational inertia, and organizational mediocracy. To mitigate these dynamics, we discuss the implications of user research for red teaming and the importance of embedding red teaming throughout the entire development cycle of GenAI systems.}
\end{abstract}



\received{October 2024}
\received[revised]{April 2025}
\received[accepted]{August 2025}

\maketitle

\section{Introduction}

Red teaming for generative AI (GenAI) is a method for uncovering potential harms in AI products or systems by deliberately provoking failures through adversarial attacks or adversarial prompts ~\cite{inie2025summon}. Framed as a preemptive measure to improve GenAI safety and protect end users, red teaming is widely supported among artificial general intelligence (AGI) \footnote{According to Schuett et al. ~\cite{schuett2023towards}, artificial general intelligence (AGI) refers to ``AI systems that achieve or exceed human performance across a wide range of cognitive tasks.'' AGI aims to develop general AI systems capable of generalizing beyond its original training objectives and executing tasks it was not explicitly designed for ~\cite{ramanNavigatingArtificialGeneral2025}. While AGI and generative AI (GenAI) are distinct—AGI referring to broad cognitive capabilities and GenAI emphasizes content generation—the terms can overlap in practice.}researchers. In one survey, 98\% of leading AGI experts somewhat or strongly agreed that AGI labs should conduct pre-deployment risk assessments, evaluate models for dangerous capabilities, commission third-party model audits, establish safety restrictions on model usage, and engage external red teams ~\cite{schuett2023towards}. 

As GenAI rapidly advances ~\cite{bengioManagingExtremeAI2024, AISafetyETO2024} and becomes integrated into various scenarios of daily life, red teaming efforts have lagged behind in addressing the growing risks posed to users and must now catch up with the expanding scope of generative AI applications ~\cite{tonerExploringClustersResearch2022, AISafetyETO2024}. These applications now extend across a wide range of specialized contexts and user groups, including students learning human-computer interactions in higher education ~\cite{kharrufaPotentialImplicationsGenerative2024, bullGenerativeArtificialIntelligence2024}, K–12 students ~\cite{bryantArtificialIntelligenceEducation2020, klopferGenerativeAIK122024}, medical professionals performing clinical tasks ~\cite{melvinAIPoweredWebApplications2024, manuboluGenerativeAIDisease2024, umetonGPT4CancerCenter2023a}, and individuals undergoing psychological care ~\cite{abramsWhatPsychologistsNeed2024}, etc. These developments extend GenAI safety concerns beyond general-purpose chatbots (e.g., ChatGPT) and highlight the need to address risks emerging in situated and user-centered contexts.

In recent policy discussions and corporate initiatives, red teaming has emerged as a critical bulwark to safeguard users against the increasing risks associated with AI technologies ~\cite{lucaj2023ai, ganguli2022red, ganguli2022predictability}. Streamlined red teaming has been adopted by organizations to ensure AI safety, including large technology companies (e.g., Microsoft ~\cite{kumarMicrosoftAIRed2023}, Google \cite{fabianGooglesAIRed2023}, OpenAI \cite{openaiGPT4TechnicalReport2024}) and government agencies ~\cite{lucaj2023ai, ganguli2022red, ganguli2022predictability}, often operating under the assumption that red teaming is a one-size-fits-all solution to AI safety. However, limited attention has been paid to the harms that emerge from user behaviors in specific contexts \cite{dominique2024prompt, li2024hunyuan, ganguli2022red, ganguli2022predictability, xu2021bot}, the ways in which risks vary across user groups with heterogeneous values and ethical frameworks \cite{kirk2023personalisation, feffer2024red, ganguli2022predictability}, and the organizational practices of red teamers responsible for identifying and mitigating these risks.

At the same time, organizational dynamics can impact, distort or reduce the effectiveness of red teaming \cite{ackermanRedTeamingCrisis2021, seidCyberResilienceUsing2025}. Research on responsible AI has shown that such organizational dynamics shape how resources are allocated, how ethics are scoped and operationalized, and how the roles and identities of practitioners are constructed \cite{widderPowerPlayInvestigating2024, wangDesigningResponsibleAI2023, deng2023understanding}. Yet, compared to technical aspects, the influence of organizational dynamics on red teaming for GenAI has received little attention in HCI and CSCW. In response, we adopt an organizational perspective to examine how red teaming is practiced across different organizations. We conducted interviews with 17 practitioners—including GenAI system red teamers, GenAI content moderators, and GenAI product managers—across various organizations. Specifically, we probed the following three research questions:

(1)	\textit{What does daily red teaming work look like in organizations?}

(2)	\textit{How are red teaming strategies, automated red teaming tools, and risk assessment framework developed?}


(3)	\textit{What are the organizational dynamics influencing the red teaming implementation, and how do they shape its effectiveness in practice?}





This paper makes three key contributions. First, we provide an underexamined account of how red teaming workflows are enacted within real-world organizational settings, particularly those developing GenAI systems. Second, we offer an empirical analysis of the organizational dynamics that shape the effectiveness of red teaming implementation—specifically, organizational resistance, organizational inertia, and organizational mediocracy. Third, we discuss how user research and the integration of red teaming throughout the GenAI development lifecycle can help mitigate these organizational challenges. Taken together, these contributions move beyond technical perspectives to offer a social, political, and organizational understanding of red teaming in the context of GenAI systems.

\section{Related Work}
As we study red teaming practices in organizations developing GenAI systems, we review prior research in two domains: red teaming practices, particularly in the context of GenAI and how Responsible AI (RAI) practitioners perceive and address AI-related harms within organizational settings. This review provides the foundation for our empirical analysis of how red teamers navigate these practices in real-world organizational contexts.

\subsection{Red Teaming in the Context of Generative AI}
Red teaming, originated in military contexts, traditionally involves simulating adversary tactics to evaluate and improve strategic plans ~\cite{longbine2008red}. Over time, this approach expanded beyond military settings into domains such as cybersecurity and business competition, where red teaming is used to discover vulnerabilities, challenge assumptions, and strengthen decision-making processes by anticipating adversarial behaviors ~\cite{5692200, zenko2015red}. In the context of generative AI, red teaming involves adversarial testing to identify cyber vulnerabilities (e.g., testing cyber-resiliency under degraded conditions) ~\cite{randhawa2018mission, abbass2011computational, li2022collaborating, al2018friendly, wang2017contrasting}, conducting algorithm audits (e.g., detecting discriminatory classification performance) ~\cite{bandy2021problematic}, and content moderation (e.g., identifying harmful outputs generated by GenAI) ~\cite{ganguli2022red, kirk2023personalisation}, making it a complex cross-disciplinary activity ~\cite{kirk2023personalisation, hung2022arl, chang2024red}. Existing research has noted several unresolved issues in red teaming for GenAI: First, model alignment strategies, such as using reinforcement learning with human feedback (RLHF) to refine the moral judgments of GenAI systems, cannot fully prevent malicious manipulation from users ~\cite{narayananModelAlignmentProtects2023a, ganguli2022predictability}. Second, how red teaming should be implemented in industry and what the process should entail remains an open question. As Feffer et al. ~\cite{feffer2024red} noted, the absence of clear guidelines and reporting standards undermines red teaming efforts. To resolve this, organizations need to define the goals of red teaming, align them with practical threat models, and establish concrete follow-up actions such as reporting, disclosure, and mitigation ~\cite{feffer2024red}. Third, collaboration among domain experts, automated red teaming tools, AI red teamers, and users has not been fully examined in workplace settings. As AI models’ interaction capabilities advance, the boundaries between conventional red teaming, algorithm auditing, and content moderation have become increasingly blurred. This blurring complicates the structure of human labor and heightens the need for shared responsibility among AI models, developers, domain experts, and field workers ~\cite{sambasivan2022deskilling}. However, Sambasivan and Veeraraghavan ~\cite{sambasivan2022deskilling} found that domain experts and field workers are often relegated to data collection roles, with little involvement in system design or decision-making processes.

The current body of research on red teaming methodologies has primarily focused on the generative models. Humphreys et al. identified four major vulnerabilities in GenAI: data poisoning, training data extraction, backdooring, and adversarial prompting ~\cite{humphreys2024ai}. Compared with red teaming for earlier AI systems, adversarial prompting is a distinct approach to GenAI and includes techniques such as prompt injection and adversarial examples ~\cite{LargeModelSecurity2024}. To complement the limitations of manual red teaming, such as the inability to test every topic of interest ~\cite{dominique2024prompt}, some companies have adopted automated tools, example databases, or prompt templates. Automated tools typically perform two key tasks: generating attack samples and conducting risk analysis ~\cite{li2024hunyuan}. For example, Tencent’s Hunyuan model is a text-to-image diffusion transformer designed with fine-grained comprehension of both Chinese and English, designed to ensure compliance with various laws and regulatory attempts, such as the ``Interim Measures for the Management of Generative AI Services'' ~\cite{li2024hunyuan}. Effective manual red teaming, meanwhile, requires experts to have both domain-specific knowledge and a deep understanding of how to probe AI models for unexpected behaviors ~\cite{ganguli2022red}. Prior research has also explored semi-automated techniques that incorporate human feedback as an intervention, aiming to combine the strengths and mitigate the limitations of both fully automated and fully manual red teaming approaches ~\cite{xu2021bot}. 

Despite these methodological leaps, less attention has been given to defining user groups and assessing whether their values align with existing red teaming methodologies. Additionally, issues such as ambiguous implementation processes and limited collaboration among domain experts, automated red teaming tools, AI red teamers, and users remain unresolved. Our study explores these gaps by investigating how red teamers engage with these approaches in practice and how they contend with the related situational challenges that arise from their daily red teaming work.


\subsection{The Heterogeneity of Perceived Harm Upon Users}

To uncover harms in GenAI systems, red teaming often involves risk taxonomies and threat models to guide what types of harms to look for, which user groups to consider, and how adversarial scenarios are prioritized and evaluated. GenAI risk taxonomies primarily focus on concerns in AI-generated outputs such as mis/disinformation, harmful instructions, and bias toward vulnerable groups ~\cite{koesslerRiskAssessmentAGI2023, ganguli2022red, dominique2024prompt, sison2024chatgpt, nagireddy2024socialstigmaqa, weidinger2021ethical}. Threat models, in turn, articulate assumptions about potential adversarial actors, goals, and system vulnerabilities. Feffer et al. ~\cite{feffer2024red} propose categorizing red teaming threat models according to their clarity and consensus: dissentive (unclear risks), consentive (well-defined risks), both, and risks formulated from scratch. 

When interactions from non-malicious users are taken into account, developing effective taxonomies and threat models to surface the risks before deployment becomes more complex due to the unpredictable and context-dependent nature of AI-generated outputs ~\cite{mehrabi2023flirt, dorotic2024ai}. Recognizing the limits of purely technical solutions, responsible AI (RAI) scholars increasingly emphasize the need to involve end users and domain stakeholders throughout the AI development lifecycle, particularly in defining fairness, safety, and harm. Smith et al. ~\cite{Smith2023Scoping} show that co-designing fairness metrics with stakeholders such as content providers on AI-driven platforms leads to metrics that better reflect lived experiences of unfairness and context-specific needs, bridging a gap often missed when developers act alone. Deng et al. ~\cite{Deng2022ExploringHow} found that industry practitioners struggled to apply one-size-fits-all ``fairness toolkits'' to their own AI projects. This suggests that these threat models need to be refined through direct participation of practitioners to fit real-world use cases. Yildirim et al. ~\cite{Yildirim2023Investigating} further highlight how practitioners use human-AI guidelines such as the People + AI Guidebook not only to inform design decisions but also to educate colleagues and collaborate across teams, implying that practitioners’ feedback is essential in evolving these guidelines to address real-world challenges.

Researchers also found that the operational work of Responsible AI often falls disproportionately on marginalized individuals who take on these tasks as volunteer efforts that rarely advance their careers ~\cite{Ali2023Walking, Deng2022ExploringHow}. Widder et al. ~\cite{widderPowerPlayInvestigating2024} describe how AI ethics duties are treated as low-status ``chores,'' frequently assigned to women or other underrepresented team members rather than distributed evenly. This uneven distribution of labor is exacerbated by a lack of effective cross-functional collaboration processes, with industry professionals often having to engage in ``bridging work'' across roles to overcome frictions in understanding and evaluating AI fairness issues ~\cite{dengInvestigatingPracticesOpportunities2023}. RAI tasks such as fairness reviews, bias audits, and harm assessments often constitute invisible labor that goes unrecognized by managers, leaving practitioners from minority groups feeling undervalued and disproportionately burdened, often leading to emotional strain and burnout ~\cite{Deng2022ExploringHow}. Acknowledging and supporting these contributors is crucial because their lived experiences and advocacy often surface harms ~\cite{cheon2024amazon,cheon2024creative} that would otherwise be overlooked.

Collectively, these findings suggest that proactive user and stakeholder engagement through co-design workshops and practitioner-centered tool development can help account for the heterogeneity in perceptions of harm and better align GenAI systems with the values and expectations of diverse users. At the same time, fully realizing participatory RAI also requires supporting the marginalized workers who carry out much of this work. Hence, embracing these participatory approaches in GenAI development and providing institutional support for the people doing this invisible labor is a promising strategy to anticipate and mitigate a broader range of harms. This paper examines how these heterogeneous perceptions of harm manifest in actual red teaming practices within organizations. While previous work has focused on developing theoretical frameworks for user involvement, we empirically investigate how red teamers navigate these varying definitions of harm in practical settings with insufficient guidance or user research. 



\subsection{Red Teaming as Organizational Practices}


Organizational factors critically shape the effectiveness of red teaming and broader responsible AI (RAI) efforts. As researchers benchmark GenAI systems against human-value criteria through risk audits, safety scorecards, and ethics analyses ~\cite{oneilAuditingAlgorithmicRisk2024, jabbourGenerativeAIAgents2024, iyerLightShadowsGenerative2024}, evaluation results and metrics may still fail to capture qualitative error patterns or address critical conceptual flaws. These benchmark performances can also vary significantly depending on evaluation context and prompting strategies ~\cite{roy2024beyond, jinhidden, tao2024gpt, madaio2022assessing}. These limitations reflect a pattern of means-ends decoupling, in which well-intentioned practices fall short of achieving their intended outcomes ~\cite{Bromley01062012}. As organizational theory suggests, when RAI practices are difficult to evaluate and the relationship between means and ends is opaque, organizations tend to develop increasingly complex internal structures, persist in perpetual reform cycles, and divert resources away from core goals ~\cite{Bromley01062012, Ali2023Walking}. For red teaming, this means efforts may become procedural formalities rather than effective safeguards, with resources spent on process compliance instead of uncovering and addressing real risks. Organizations also face strong counter-incentives to disclose red teaming results due to potential reputational risks ~\cite{ganguli2022predictability}, creating policy-practice decoupling where organizations may adopt AI safety policies without implementing corresponding practices ~\cite{Bromley01062012}. RAI researchers have observed similar patterns of decoupling in organizational ethics efforts, where policies, practices, and outcomes diverge, and collective action within organizations is often absent ~\cite{Ali2023Walking}. Ali et al. ~\cite{Ali2023Walking} identify three core challenges to re-couple policies, practices, and outcomes in organizational RAI practices: (1) prioritizing ethics under product launch pressures; (2) quantifying ethics in metrics-focused environments; and (3) team reorganizations that disrupt institutional memory and key relationships. These dynamics often individualize risk, making the act of raising ethical concerns a personal liability—particularly for workers from marginalized backgrounds.

Recent work suggests that addressing these challenges requires mobilizing diverse stakeholders and adopting strategies that integrate ethics into everyday workflows. Liao and Xiao ~\cite{liao2023rethinking} point to two dimensions for improving GenAI safety evaluations: context realism and human requirement realism. Dominique et al. ~\cite{dominique2024prompt} show that prompt templates can help outside participants, also known as non-expert red teamers, supplement their red teaming knowledge and improve their red teaming performance. Practitioners have also developed various strategies to confront organizational resistance to AI ethics work, such as the ``soft resistance'' of UX professionals—making values visible, creating space for ethical reflection, and gradually shifting the workflows ~\cite{Wong2021Tactics}. Similarly, Deng et al. ~\cite{dengInvestigatingPracticesOpportunities2023} find that fairness considerations can be embedded into existing requirements such as privacy assessments. Wang et al. ~\cite{wangDesigningResponsibleAI2023} describe three emerging strategies: building RAI lenses, responsible prototyping, and RAI-specific evaluations. Rakova et al. ~\cite{Rakova2021WhereRAI} further emphasize that aligning organizational structures with daily practices is essential to translating responsible AI policies into effective implementation.

Despite these insights, few studies have stretched their observations to organizational dynamics in the context of red teaming, leaving a gap in systematic examinations through the lens of organizational factors ~\cite{feffer2024red, madaio2022assessing, dominique2024prompt, liao2023rethinking}. Understanding organizational red teaming requires attention to the intersections between red teaming as a technical practice, organizational goals, and structural constraints  ~\cite{madaio2022assessing, dominique2024prompt, liao2023rethinking}. Our study addresses this gap by investigating how red teaming is implemented across diverse organizational contexts and how organizational dynamics shape both the definition and execution of AI safety measures. Drawing on the lived experiences of red teamers, we show how resource allocation, task distribution, and competing priorities influence the effectiveness of red teaming, offering a grounded view of how organizational structures shape AI safety outcomes. 


\section{Method}

\subsection{Study Design and Recruitment}
To investigate our research questions about the daily work of red teaming, red teaming strategies, and risk assessment framework, we conducted a qualitative study via semi-structured interviews with 17 participants involved in or closely related to red teaming across various sectors in China. Participants were initially recruited via professional and social networking platforms, including LinkedIn, X, WeChat, and Discord, followed by snowball sampling, wherein initial participants were asked to help us recruit additional red teamers from their networks. These interviews were conducted online, with all participants providing both oral consent during the session and a signed consent form as required by the IRB approval obtained from the authors' university. All participants were pre-screened to ensure they were over 18 years old and actively involved in or closely related to red teaming activities specific to large-scale generative AI models, including adversarial attack, adversarial testing, and harmful content moderation. All participants also had direct experience handling harmful content generated by AI. To ensure relevance to our research focus, we excluded participants whose red teaming work did not involve harms on users. For example, two participants who conducted red teaming on AI systems in automated textile manufacturing were excluded, as their work focused on identifying risks related to textile production (e.g., substandard textile products) rather than assessing the potential impact of AI-generated content on users.

\subsubsection{Participants} 
The participants had diverse professional backgrounds, including internal red teamers, outsourced red teamers, engineers performing red teaming tasks, product managers responsible for red teaming activities, dedicated red teaming researchers, and one domain expert on risk and crisis management. Participants' ages ranged from 20 to 35, with an average age of 26 years. The gender ratio shows a strong disparity with only two female participants, reflecting the broader gender imbalance in China’s tech industry and the underrepresentation of women in technical roles ~\cite{yuanChinaTechWorkers2017}. 

Participants’ red teaming experience ranged from under a year—among more than half of them—to up to four years. Their work spanned diverse generative AI models, testing scopes, and organizational settings. Most of the participants (12 out of 15) reported working primarily with text-based generative AI models. Image-based models followed, with eight participants engaged in their risk assessment. Video and audio models represented a smaller focus, with only two participants each testing the risks in video-generative or audio-generative models. Additionally, two respondents noted that their testing scope depended on specific client needs or internal organizational tasks (e.g. adversarial tests against misinformation dissemination for client's GenAI bot on social media). In terms of organizational characteristics, two participants work in third-party independent red teaming companies, two participants worked for government projects as interns or external domain experts, five are red teaming researchers in universities, and the rest six are internal organization red teamers or engineers/managers responsible for AI risks at tech companies. Additional participant details are provided in Table 1.

\begin{table}[]
\centering
\resizebox{\textwidth}{!}{%
\begin{tabular}{c|c|c|c|c|c|c}
\hline
\textbf{\begin{tabular}[c]{@{}c@{}}Participant \\ ID\end{tabular}} &
  \textbf{Gender} &
  \textbf{Age} &
  \textbf{\begin{tabular}[c]{@{}c@{}}Role \\ as a Red Teamer\end{tabular}} &
  \textbf{\begin{tabular}[c]{@{}c@{}}Red Team \\ Size\end{tabular}} &
  \textbf{\begin{tabular}[c]{@{}c@{}}Affiliated \\ Organization\end{tabular}} &
  \textbf{\begin{tabular}[c]{@{}c@{}}Red Teaming \\ Field\end{tabular}} \\ \hline
\textbf{1} &
  Male &
  19 &
  Internal red teamer &
  8 &
  Tech company &
  \begin{tabular}[c]{@{}c@{}}Race and \\ religion bias\end{tabular} \\ \hline
\textbf{2} &
  Male &
  20 &
  Internal red teamer &
  8 &
  Tech company &
  \begin{tabular}[c]{@{}c@{}}Race and \\ religion bias\end{tabular} \\ \hline
\textbf{3} &
  Female &
  27 &
  Internal red teamer &
  10 &
  Tech company &
  E-commerce \\ \hline
\textbf{4} &
  Male &
  20 &
  \begin{tabular}[c]{@{}c@{}}Internal engineer \\ doubling \\ as a red teamer\end{tabular} &
  4 &
  \begin{tabular}[c]{@{}c@{}}Government \\ research institute\\  \& University\end{tabular} &
  \begin{tabular}[c]{@{}c@{}}Geology \\ \& Computer \\ vision\end{tabular} \\ \hline
\textbf{5} &
  Male &
  22 &
  \begin{tabular}[c]{@{}c@{}}Red teaming \\ researcher\end{tabular} &
  NA &
  University &
  Not specific \\ \hline
\textbf{6} &
  Male &
  35 &
  \begin{tabular}[c]{@{}c@{}}Product manager \\ with red teaming \\ responsibilities\end{tabular} &
  NA &
  Tech company &
  \begin{tabular}[c]{@{}c@{}}K-12 \\ education\end{tabular} \\ \hline
\textbf{7} &
  Male &
  28 &
  Outsourced red teamer &
  \begin{tabular}[c]{@{}c@{}}Depends \\ on project\end{tabular} &
  Tech company &
  \begin{tabular}[c]{@{}c@{}}Not specific, \\ depends on \\ what project \\ they have got\end{tabular} \\ \hline
\textbf{8} &
  Male &
  22 &
  Outsourced red teamer &
  6 &
  Tech company &
  \begin{tabular}[c]{@{}c@{}}Depends \\ on the client\end{tabular} \\ \hline
\textbf{9} &
  Male &
  27 &
  Internal red teamer &
  15 &
  Tech company &
  \begin{tabular}[c]{@{}c@{}}Depends \\ on the task\end{tabular} \\ \hline
\textbf{10} &
  Male &
  24 &
  Internal red teamer &
  50 &
  Tech company &
  Social media \\ \hline
\textbf{11} &
  Female &
  26 &
  \begin{tabular}[c]{@{}c@{}}Internal engineer \\ doubling \\ as a red teamer\end{tabular} &
  10 &
  Tech company &
  Fashion \\ \hline
\textbf{12} &
  Male &
  28 &
  Red teaming researcher &
  NA &
  University &
  Not specific \\ \hline
\textbf{13} &
  Male &
  30 &
  External expert &
  10 &
  \begin{tabular}[c]{@{}c@{}}University \\ \& Government\end{tabular} &
  Not specific \\ \hline
\textbf{14} &
  Male &
  28 &
  \begin{tabular}[c]{@{}c@{}}Internal engineer \\ doubling \\ as a red teamer\end{tabular} &
  20-30 &
  Tech company &
  \begin{tabular}[c]{@{}c@{}}Psychological \\ therapy\end{tabular} \\ \hline
\textbf{15} &
  Male &
  21 &
  Red teaming researcher &
  8 &
  University &
  Not specific \\ \hline
\textbf{16} &
  Male &
  27 &
  Red teaming researcher &
  7-8 &
  University &
  Not specific \\ \hline
\textbf{17} &
  Male &
  25 &
  Red teaming researcher &
  10 &
  University &
  Not specific \\ \hline
\end{tabular}%
}
\caption{The participants' demographic information and their red teaming fields (NA means not applicable)}
\label{tab:participant_info}
\end{table}

\subsection{Interview Protocol}
We developed our semi-structured interview protocol iteratively through the first two interviews (P1, P2). Following these interviews, we reviewed and refined some questions for better clarity, and their sequence was optimized to enhance conversational flow. The interviews for the remaining 15 participants who took part in the formal study lasted an average of 46 minutes, with a range of 26 to 102 minutes. The same protocol was used for both industry and academic red teamers to ensure comparability across participant responses. For example, we first asked each participant to introduce themselves and describe their experience with generative AI (e.g., ``How long have you been using generative AI, and what are your overall impressions of it?''). Participants then described their specific roles, responsibilities, and typical workflows (e.g., ``Could you briefly describe your position and your typical day at work?''). Next, we focused on their understanding and practical approaches to red teaming (e.g., ``How would you describe red teaming work? Can you explain your strategies for testing and mitigating harmful AI-generated content?''). We explored their experiences and the challenges encountered while doing the red teaming related works (e.g., ``What types of harmful content do you typically test for, and what challenges have you faced?''). Participants also discussed their emotional reactions to encountering harmful content (e.g., ``How do you typically respond to harmful AI-generated content? Could you provide examples?''). We further explored the use and perceptions of automated tools and databases (e.g., ``Do you use any automated tools or databases to assist with your red teaming tasks? How helpful are they compared to human team members?'') and the workplace dynamics~\cite{cheon2022dynamic,workgame2025} within their organizations (e.g., ``How do you work with colleagues or departments inside and outside your organization before and after identifying risks?''). Finally, we addressed stress management and workplace support mechanisms (e.g., ``How does your organization protect your mental health when exposed to harmful content? What practices are employed by leadership to support your well-being?''). Participants were also encouraged to envision and describe their ideal red teaming work environment and workflow.

\subsection{Data Analysis}

Following the completion of the interviews, we conducted a thematic analysis to identify key patterns and recurring themes in participants’ experiences with red teaming in generative AI systems. Our analysis followed an inductive approach grounded in open coding, allowing us to remain attentive to unexpected insights and emergent tensions across the dataset ~\cite{saldanaCodingManualQualitative2015a, auerbachQualitativeDataIntroduction2003}. Two coders independently reviewed five to ten transcripts during the initial round of open coding, generating hundreds of descriptive and interpretive codes. Throughout this process, coders wrote analytic memos, documented reflections, and raised conceptual questions. Discrepancies in coding were discussed collaboratively, and codes were iteratively refined through team discussions and negotiated consensus.

To support the structuring and interpretation of our findings, we drew on the \textit{Hierarchy of Influences model} ~\cite{shoemakerMediatingMessage21st2013a}, which was originally developed in media studies to conceptualize the multilayered influences shaping content production. The model includes five levels of influence: individual, routine practices, organizational structures, social institutions, and societal systems ~\cite{shoemakerMediatingMessage21st2013a}. While we did not apply this framework deductively, we adapted it to reflect the domains most relevant to our research questions. Specifically, we focused on three levels—individual, organizational, and technological—as these emerged consistently across participants' accounts and offered a lens for organizing the dynamics at play in red teaming work.

Based on this framework and the patterns identified through open coding, we developed two coding schemes. The first focused on the broader contextual factors shaping red teaming work, capturing how individual roles, organizational practices, and technological infrastructures influenced daily workflows (see Table 2). The second coding scheme addressed the specific practices of red teaming implementation, including strategies, automated tools and datasets, and the criteria used to assess AI risks (see Table 3). We then integrated these two coding schemes into a matrix that allowed us to analyze how different levels of influence intersected in practice. In total, we generated 266 unique codes, which we grouped into 31 categories (e.g., Criteria for Red Teaming → Identifying AI Risks → Individual Factors). These categories were subsequently distilled into higher-order themes that reflect shared challenges, variations in practice, and organizational tensions in how red teaming is operationalized within GenAI development settings. These themes form the basis of the findings presented in the following sections.

\begin{table}[]
\centering
\resizebox{\columnwidth}{!}{%
\begin{tabular}{c|l|l}
\hline
\textbf{Category} & \multicolumn{1}{c|}{\textbf{Definition}} & \multicolumn{1}{c}{\textbf{Anchor Example}} \\ \hline
\textbf{\begin{tabular}[c]{@{}c@{}}Organizational \\ factors\end{tabular}} &
  \begin{tabular}[c]{@{}l@{}}These relate to the internal dynamics, goals, and structures \\ within organizations that influence the adoption \\ and execution of red teaming practices. This category \\ also includes variations across industries that might affect \\ how red teaming is prioritized and operationalized.\end{tabular} &
  \begin{tabular}[c]{@{}l@{}}``Initially, when we generated fabricated weather \\ disaster information, we added entertaining elements \\ to make it engaging. After reviewing it, leadership \\ thought this could work for social media promotion \\ to drive traffic to our product. {[}...{]} But eventually, \\ they ordered us to halt the attacks—they were furious. \\ They want the blue team to win.'' (P8)\end{tabular} \\ \hline
\textbf{\begin{tabular}[c]{@{}c@{}}Technology \\ factors\end{tabular}} &
  \begin{tabular}[c]{@{}l@{}}These pertain to non-human, technical aspects of \\ generative AI and red teaming, focusing specifically \\ on algorithms, mechanisms, and tools. They deal with \\ the inherent technological challenges and capabilities \\ related to AI models, such as their alignment, robustness, \\ and security vulnerabilities.\end{tabular} &
  \begin{tabular}[c]{@{}l@{}}``In white box red teaming, I use the discrete optimization\\ algorithms to generate red teaming prompts as well as \\ samples to test the text-to-text model. For black-box, I \\ will use zero-order methods or specific techniques to \\ generate the objects.'' (P12)\end{tabular} \\ \hline
\textbf{\begin{tabular}[c]{@{}c@{}}Individual \\ factors\end{tabular}} &
  \begin{tabular}[c]{@{}l@{}}These are personal characteristics and demographic \\ elements that influence one’s engagement with or experience \\ in red teaming and AI use. This includes factors like gender, \\ age, personal relationships, and other individual \\ demographic features that affect participation, \\ perspectives, or task allocation in the AI red teaming \\ ecosystem.\end{tabular} &
  \begin{tabular}[c]{@{}l@{}}``When male colleagues presented something related \\ to pornography detection or identification, {[}…{]} \\ If a male colleague were the one showing it \\ (to red teamers), it would feel slightly uncomfortable \\ {[}…{]}. However, if a female colleague presented the \\ same material, the atmosphere in the room would \\ be much more normal.'' (P3)\end{tabular} \\ \hline
\end{tabular}%
}
\caption{Coding Scheme for Organizational, Technological, and Individual Factors. These categories are not mutually exclusive. For example, dynamics related to self-identity are shaped by both organizational factors and individual factors.}
\label{tab:my-table}
\end{table}

\begin{table}[]
\centering
\resizebox{\columnwidth}{!}{%
\begin{tabular}{l|l|l|l}
\hline
\textbf{Category} &
  \textbf{Sub-category} &
  \textbf{Definition} &
  \textbf{Anchor Example} \\ \hline
\multirow{3}{*}{\begin{tabular}[c]{@{}l@{}}Implementation \\ of Red Teaming\end{tabular}} &
  pre-activity &
  \begin{tabular}[c]{@{}l@{}}The planning and preparation phase \\ prior to the actual red teaming exercise.\end{tabular} &
  \begin{tabular}[c]{@{}l@{}}Creating a risk assessment matrix to \\ prioritize which systems to test based \\ on vulnerability and impact.\end{tabular} \\ \cline{2-4} 
 &
  during activity &
  \begin{tabular}[c]{@{}l@{}}The operational phase where red \\ teaming tasks are executed.\end{tabular} &
  \begin{tabular}[c]{@{}l@{}}Attempting to penetrate network \\ defenses using simulated attacks to \\ assess the effectiveness of security \\ protocols.\end{tabular} \\ \cline{2-4} 
 &
  post-activity &
  \begin{tabular}[c]{@{}l@{}}The follow-up actions after a red \\ teaming exercise.\end{tabular} &
  \begin{tabular}[c]{@{}l@{}}Compiling a report detailing the \\ vulnerabilities discovered during testing \\ and proposing mitigation strategies.\end{tabular} \\ \hline
\multirow{3}{*}{\begin{tabular}[c]{@{}l@{}}Strategies, \\ Automated Tools, \\ and Databases\end{tabular}} &
  automated tools &
  \begin{tabular}[c]{@{}l@{}}Software applications and systems used \\ to automate certain aspects of red \\ teaming exercises.\end{tabular} &
  \begin{tabular}[c]{@{}l@{}}Using an automated penetration testing \\ tool to systematically identify and \\ exploit weaknesses.\end{tabular} \\ \cline{2-4} 
 &
  \begin{tabular}[c]{@{}l@{}}database and \\ prompt templates\end{tabular} &
  \begin{tabular}[c]{@{}l@{}}Pre-configured data sets and scripted \\ prompts that facilitate standardized \\ testing and scenario modeling during \\ red team exercises.\end{tabular} &
  \begin{tabular}[c]{@{}l@{}}Utilizing a database of common attack \\ vectors to test system responses under\\ controlled, simulated conditions.\end{tabular} \\ \cline{2-4} 
 &
  strategies &
  \begin{tabular}[c]{@{}l@{}}Overall plans and methodologies \\ adopted for red teaming to effectively \\ identify, analyze, and mitigate potential \\ security threats.\end{tabular} &
  \begin{tabular}[c]{@{}l@{}}Developing a multi-layered attack \\ scenario to evaluate the depth of \\ defense mechanisms.\end{tabular} \\ \hline
\multirow{3}{*}{\begin{tabular}[c]{@{}l@{}}Criteria of \\ Red Teaming\end{tabular}} &
  \begin{tabular}[c]{@{}l@{}}goals of red \\ teaming\end{tabular} &
  \begin{tabular}[c]{@{}l@{}}The specific objectives set out to \\ achieve through red teaming.\end{tabular} &
  \begin{tabular}[c]{@{}l@{}}The goal is to enhance user experience \\ when using the GenAI application.\end{tabular} \\ \cline{2-4} 
 &
  \begin{tabular}[c]{@{}l@{}}red teaming \\ for whom\end{tabular} &
  \begin{tabular}[c]{@{}l@{}}Identifying the stakeholders or target \\ groups for whom red teaming is \\ conducted.\end{tabular} &
  \begin{tabular}[c]{@{}l@{}}Clear understanding of whom the red \\ teamers are protecting in red teaming\\ exercises.\end{tabular} \\ \cline{2-4} 
 &
  what is AI risks &
  \begin{tabular}[c]{@{}l@{}}Specific threats and vulnerabilities \\ associated with artificial intelligence \\ systems that red team aims to mitigate.\end{tabular} &
  \begin{tabular}[c]{@{}l@{}}Testing AI-driven decision-making \\ systems for bias under scenarios of \\ manipulated data inputs to ensure \\ fairness and accuracy.\end{tabular} \\ \hline
\end{tabular}%
}
\caption{Coding Scheme for Implementation of Red Teaming, Strategies, Automated Tools and Databases, And Criteria of Red Teaming. These categories are not mutually exclusive.}
\label{tab:my-table}
\end{table}
\section{Findings}

We begin by outlining how red teaming work is embedded within everyday routines of organizations. We then examine three key organizational dynamics—resistance, inertia, and mediocrity—that shape how red teaming is implemented. These dynamics influence not only the selection of red teaming strategies and targets but also the allocation of resources and the perceived legitimacy of the work itself. Finally, we show how red teaming becomes a form of marginalized labor, disproportionately assigned to and affecting marginalized groups—particularly in the context of moderating harmful content and in settings where organizational support is lacking.



\subsection{Daily Work as Red Teamers}

In this section, we describe the everyday work of red teamers across organizations, focusing on how their roles are defined and enacted in practice. We begin by outlining how red teaming is scoped and defined, and how it differs across academic and industry settings. We then describe how boundaries are drawn between red teams and blue teams. Then, we describe how user research is overlooked or deprioritized in current red teaming practices, and examine how automated tools are being integrated into red teaming workflows.



\subsubsection{What Counts as Red Teaming in Generative AI?}

Participants in our study offered nuanced and at times divergent understandings of what constitutes ``red teaming,'' shaped by their organizational contexts and the specific systems under scrutiny. For instance, participants working at red teaming outsourcing firms (P7, P8)—who were contracted to evaluate generative AI systems developed externally—positioned red teaming as adversarial simulation oriented toward technical risk detection. Their accounts framed red teaming primarily as a security-driven task, emphasizing the role of red teamers as stand-ins for malicious actors tasked with surfacing system-level vulnerabilities:\textit{``simulating adversarial hackers to discover the problems in networks or systems, including front-end security bugs and other security flaws''} (P7).

In contrast, participants employed by technology companies described red teaming as a more interpretive and evaluative process, oriented toward assessing system responses and human factors. As P9 explained:\textit{``I'm not here just to attack or steal information. I'm here to test the systems and users - to evaluate their reactions and defense measures. […] So I believe I'm not using this technology to steal information or do anything illegal. This is why I find red teaming work very meaningful.''} (P9)


Several participants drew comparisons between conventional red teaming and the distinct challenges posed by red teaming generative AI systems. One participant explained: 
\begin{quote}
\textit{``Traditional red teaming requires more physical and hardware-level expertise, whereas now red teaming primarily relies on intelligence and computing power. Now the harm, as harmful content, has become more subtle and implicit, not directly harmful. It requires adding specialized model training with adversarial content to complete effective attacks.''} (P15)
\end{quote}

This shift reflects the architectural and operational differences posed by generative AI. As P7 noted, the partial openness of generative models introduces new vectors of attack: \textit{``Another thing is data attacking, because generative AI systems are half open-source and cannot completely be closed-loop''} (P7). Participants (P3, 5, 10, 13, 15) also described how red teaming for generative AI models emphasizes preemptively identifying harmful content generation - an area that is not always foregrounded in other types of systems. For instance, P13, a risk management expert, likened this role to proactive moderation: \textit{``Red teaming is testing in advance to avoid AI output harmful content. I think this is similar to the content moderators on the internet.''} Likewise, P15 described red teaming as a form of protective oversight: \textit{``In my view, red teams act as security guards for generative AI, safeguarding the safety and reliability of artificial intelligence systems by preventing the harmful information they get.''}




\subsubsection{Red Teaming Strategies and Organizational Approaches}

Our participants described a range of red teaming strategies, including massive data injection (P8, P11, P7), third-person role play (P13), decontextualization or ``peeling off'' context (P12), and data contamination (P7). These strategies reflected not only individual creativity but also broader organizational priorities and constraints around red teaming processes, AI harm detection, and evaluation criteria. Within academic settings, red teaming efforts tend to be more systematic and ethically motivated. Researchers are exploring approaches like \textit{constitutional AI}—a method designed to align generative AI systems with ethical principles by encoding adaptable guidelines directly into the models. This technique leverages both supervised and reinforcement learning to allow AI systems to maintain a flexible but principled stance across contexts. As one participant explained:\textit{``In terms of advantages, it [constitutional AI] is very convenient because you can quickly modify the values [held by AI]. Legal provisions constantly change, and different user groups have varying values, making it easy to rapidly update our constitution. The flexibility to accommodate different perspectives is a key strength since it’s embedded directly into the system''} (P12).  In contrast, industry-based red teaming often centers on identifying and mitigating harms within tighter operational constraints. For instance, P10, who works at a major social media platform, oversees a generative AI system that automatically generates comments beneath users’ social media posts. Their team's work is structured around a predetermined list of keywords, which shapes and limits the scope of harm assessment: \textit{``My role is focused on oversight. Our team has four members, and each of us is responsible for monitoring a different set of keywords. After receiving the keyword list, we hold discussions to coordinate our work; however, our discussions are limited to the assigned keywords, not having a [broader] goal to discuss.''} (P10) This contrast highlights \textbf{how institutional context — academic versus corporate — shapes not only red teaming goals but also the degree of flexibility, ethical framing, and strategy that practitioners can exercise.}

\subsubsection{Blurred Boundaries between Red and Blue Teaming} 

Traditionally, red teams simulate successful attacks to reveal system vulnerabilities, while blue teams defend against these attacks and propose mitigation strategies. However, several of the organizations we examined, these roles were not clearly delineated. Red and blue team responsibilities were often integrated within the same individuals or units.

For example, P9, a safety researcher, described a workflow that blends offensive and defensive roles: \textit{``From the very beginning of a workday, I research the latest generative AI technologies and gather information on potential threats. I review materials, meet with team members to design test plans, and set up testing environments. We then simulate attack scenarios using generative AI tools and observe the system’s responses. Afterward, I analyze the results, write a report, and share issues and improvement suggestions with relevant departments.''} Here, the red team activity of probing vulnerabilities is inseparable from the blue team task of recommending mitigations. Similarly, P11’s account highlights a hybrid approach that leans toward blue teaming while still incorporating red team tactics: \textit{``Our daily work involves data monitoring and performance testing of the app...our team discusses and coordinates each day to assign specific areas for simulated attacks. For instance, one person may attack A, while I focus on B. After testing, we provide feedback, capture relevant packages, and identify where and how to fix the problems.''} These accounts reveal how red and blue team practices often coexist within the same organizational structures and even within individual workflows. Rather than operating as discrete adversarial units, team members frequently oscillate between identifying risks and implementing solutions — a merging of roles shaped by the demands that generative AI systems place on red/blue teams, such as the need for continuous adaptation, and by the practical realities of platform defense, including resource constraints and the pressures for operational efficiency.

\subsubsection{Limited Engagement with Users in GenAI Red Teaming} 

In contrast to responsible AI practices that emphasize early-stage user research to identify and taxonomize risks, we found that red teamers rarely incorporated user perspectives when assessing harms associated with generative AI. Red teaming is often conducted after a product has been developed and just before public release, making it difficult to integrate red teaming insights into either the early stages of user experience design or post-deployment evaluations.

Few red teamers reported drawing on user research to define or scope potential harms. One participant noted that his company recruits particularly aggressive or ``high-quality'' users to serve as adversarial testers: \textit{``Because attacking GenAI often require large amounts of data stacking, red teaming now involves training individuals for this work. Therefore, our company seeks out a group of `high-quality users' and positions them as potential attackers. In our red teaming tests, part of the red team is actually composed of these users, who are embedded among professional red team members and are tasked specifically with carrying out disruptive activities''} (P7).

Some participants described attempting to act as proxies for users, claiming to take a user-centered stance when probing for harms. As P15 explained, he would \textit{``try to take the user's perspective to observe, to study, and to ask various probe questions (to the GenAI model).''} However, when asked directly who they were trying to protect through red teaming, participants’ responses were vague and inconsistent. Their answers ranged from \textit{``those groups are easily to believe others, where the AI-generated mis/disinformation goes to''} (P8), to \textit{``the teenagers whose psychological status can be impacted''} (P17) , and \textit{``only consumers (of our e-commerce platform)''} (P11).  These findings suggest a disconnect between red teaming practices and grounded understandings of real-world user risks. \textbf{Without systematic user involvement, red teaming risks becoming an internal exercise in speculative harm rather than a robust strategy for surfacing actual user concerns.}

\subsubsection{Limited Adoption of Automated Red Teaming Tools} 

Despite growing interest in automation, our participants reported limited use and widespread distrust of current automated red teaming tools. Of the 15 red teamers we interviewed, 11 expressed high expectations for automated support but ultimately refrained from relying on available tools. Participants characterized existing tools as misaligned with practical needs—describing them as \textit{``not consistent with our needs''} (P10), \textit{``low performance''} (P8), and \textit{``not accurate''} (P5). P10 illustrated the limitations of these systems in high-stakes contexts: \textit{``Some content cannot be accurately judged by automated tools and still requires manual review. Harmful content often demands human thinking to detect—robots lack the emotional capacity to understand it.''}

While most participants did not actively rely on such tools, a few selectively integrated AI into specific strategies. For instance, P7, facing the challenge of injecting large volumes of misinformation to stress-test generative models, used ChatGPT to generate high-quality variants of seed data: \textit{``I used scripts to scrape misinformation from the web, but truly impactful, high-quality fake data is scarce. That’s when I realized ChatGPT could systematically rewrite a single piece of information into multiple variants while preserving quality. After proposing this method to the team, we all adopted it. Eventually, our red team began leading attack-defense exercises instead of the blue team.''} Here, generative AI functioned not as a replacement but as a means to scale strategic labor—augmenting the capacity of a smaller red team amid an organizational labor imbalance that favored the blue team.

Looking forward, participants called for tools that are not only technically robust but also designed through inclusive, domain-informed processes. For example, P12 emphasized the importance of adaptive systems rooted in human expertise: \textit{``A good automated red teaming tool should have self-adaptive capabilities to different [contexts], and developing this requires input from red teaming experts across various fields. Experts from different domains need to contribute red teaming problems. […] Therefore, I believe that many human experts must help create some seed datasets and principles, which we can then use as a foundation to train automated red teaming tools. Human experts are needed to perform the cold-start [when there is no data, experience, or existing template for a particular domain]''} (P12). This vision for automation rooted in domain expertise highlights a critical insight: effective automation in red teaming requires more than performance optimization—it demands thoughtful integration of expertise, context, and situated knowledge.


\subsection{Organizational Dynamics in Red Teaming Practices}

Ideally, effective red teaming for generative AI (GenAI) systems begins with identifying and hypothesizing potential risks before testing commences. These hypothesized harms guide the adversarial strategies employed by red teamers, who then attempt to provoke and expose them through targeted interactions with the system. Once surfaced, these harms are expected to be addressed and mitigated by the system’s developers. 


Across our interviews, we identified three recurring organizational dynamics that constrain red teaming efforts: \textit{organizational resistance, organizational mediocrity}, and \textit{organizational inertia}. These dynamics often impede the ability of red teams to fully uncover and articulate potential harms in GenAI systems. As a result, socio-technical gaps can emerge between the risks that users should be protected from and those that red teams are practically able to discover. These gaps are frequently shaped by the relationships between red teams, blue teams (defenders), domain experts, and users—relationships that are themselves structured by institutional hierarchies and the degree to which red teaming is integrated into the development process.

\subsubsection{Organizational Resistance: Red Teaming as ``the Last Pass''} 

Originally designed to simulate malicious attacks in real-world settings, red teaming is often introduced late in the development cycle—after software has been built (N=11), and occasionally just before public release (N=4). Consequently, it is frequently framed as a final safety check rather than a process embedded across the product lifecycle. Because development is typically complete before red teaming begins, any findings may necessitate costly iterations. This can lead to resistance from within the organization. As P7 recalled: \textit{``It was clear that this pressure had been building from the start. On the fifth day (of adversarial testing), my boss had already subtly hinted that we needed to let the blue team win because it would avoid a lot of potential trouble later on.''}

Positioning red teaming at the tail end of development often creates implicit boundaries on what can be challenged. Findings that point to flaws in the algorithm’s architecture—or call into question fundamental system design choices—are seen as destabilizing, as they requiring extensive rework and risk delaying the product delivery. For instance, in a red teaming project focused on weather prediction software, P8 and colleagues discovered they could interrupt core system functionalities: \textit{``We managed to get to the backdoor and paused the weather prediction sensors that are linked to the AI system.''}. While this demonstrated a significant vulnerability, P8 explained how organizational pressures escalated once the implications became clear:
\begin{quote}
\textit{Initially, when we generated fabricated weather disaster information, we added entertaining elements to make it engaging. After reviewing it, leadership thought this could work for social media promotion. [...] But eventually, they ordered us to halt the attacks—they were furious. They want the blue team to win. [...] Later, we started receiving subtle directives from various leaders that our testing approach was unsustainable. [...] We disabled the monitoring features required for the attacks. This revealed a critical flaw: we could forcibly access backend systems, but the defense team had no means to remediate it. In other words, the system (at the architectural level) had vulnerabilities.''}
\end{quote}

A similar dynamic was described by P11, who tested a virtual try-on application for makeup and clothing. She reflected on the tension between red teamers and leadership: \textit{``We simulate scenarios where many users flood the backend, issuing different commands and tricky questions to the AI. There are also tests under weak network conditions. In the beginning, the results didn’t meet the standards required for release on the Apple App Store, which brought a lot of pressure. There's a running joke that the product director and our red team are always on opposing sides.''}

\subsubsection{Organizational Mediocracy and Regulatory Check}
When organizations take on regulatory responsibilities, their red teaming processes often default to meeting the minimum thresholds for identifying AI-related harms, as established by external authorities. This leads to a risk-averse, checkbox-driven culture of compliance.

One illustrative case involves the use of sensitive keyword lists in red teaming text-generative AI systems. These lists serve as bounded frameworks for testing, where red teamers’ activities are confined to the detection of restricted terms. For example, P10 was responsible for ensuring that their social media platform’s generative bot did not comment using banned keywords. Similarly, P13, a domain expert invited to red team a government-developed GenAI bot, described his strategy as follows: \textit{``We receive a sensitive keyword list—about four to five thousand terms. The list is compiled by government agencies and firms that monitor internet content and public sentiment. We then create prompts to see if the GenAI system reproduces any of those restricted terms.''}

On a more hopeful note, some organizations do treat red teaming as a moment of ethical responsibility, particularly when user-facing harms arise. For instance, P14 recounted their team’s response during testing of a psychological therapy AI system: \textit{``[When a user shows signs of psychological distress], we first reach out for a face-to-face conversation to offer support. Since we work with counseling institutions, we arrange for a counselor to follow up. When the user's psychological issues are more severe, the AI alone isn't enough. Therefore, we ensure a counselor steps in. We see this as our duty because users are helping us test the system.''} 

While these efforts are commendable, most compliance-centered approaches risk fostering tunnel vision: they address only harms recognized by regulators or internal priorities, leaving broader or more emergent concerns unexamined. Regulatory frameworks tend to penalize only the most egregious violations, incentivizing companies to optimize for not being the worst, rather than striving for systemic safety. This attitude was candidly articulated by P3, an engineer working on AI-generated content for an e-commerce platform
\begin{quote}
\textit{``Since safety violations are common in the industry and regulators are unlikely to single us out, we prioritize shareholder interests. We won’t be the least compliant in AI safety, but we’ll aim to be the second least. If you’re the safest, you sacrifice customer experience. Being the second least compliant, you can avoid regulatory attention (from the regulators).''}
\end{quote}
In contrast to these industry dynamics, researchers in academic or non-commercial institutions tend to engage regulators more proactively. As P15 and P16 explained, red teamers in these settings often communicate findings directly to \textit{``the regulators of these models''}—a transparency largely avoided in industry settings where the goal is to evade scrutiny rather than invite it.

\subsubsection{Inertia from Domain Familiarity: When Blue Teams and Experts Constrain Creativity}
We found that organizational inertia can emerge when red teamers work closely with blue teamers or software developers, or when domain expert perspectives substitute for those of end users.

As generative AI systems evolve, so too do the defenses that accompany them. However, familiarity with internal safeguards can inadvertently restrict red teamers’ creativity. When red teamers are internal employees, frequent collaboration and communication with blue teamers, often the same developers who trained the models, can introduce cognitive biases. P5 explained that this closeness leads to predictable attack paths and limits divergent thinking among red teamers: \textit{``Red teamers know how they (software developers as well as blue teamers) trained the model, so they will be leaning towards finding bugs from the way they train the model, getting into their defense trap and being less creative.''} Another source of inertia comes from over-focusing on known attack vectors or previously addressed vulnerabilities. P13 described that red teamers were explicitly told to avoid established strategies—even when those might still be relevant: \textit{``The manager said, ‘Use your imaginations.…If the strategy works and is already documented online, the engineers would have fixed it already’.''} 

Domain experts are frequently engaged to ensure the relevance of red teaming tasks to specific user populations. For instance, public affairs professionals are brought in to identify anti-ideology content (P13). Yet, the inclusion of experts does not always translate to user-centered insights, particularly in generative AI contexts where user expectations and interaction models differ markedly from those in conventional applications. For example, P14 reflected on efforts to design an AI-powered therapy agent, noting that while psychologists contributed during the mock-up and testing phases, their input did not fully align with how users actually interpreted generative AI behavior:\textit{``We invited some psychologists to guide AI training. They could provide suggestions, but they were not familiar with how the underlying logic and algorithms work for generative AI. The psychological scale tests have no issues by itself, but being sent from AI often let users feeling confused or dissatisfied.''} 

Unlike in traditional therapy, users of AI-based tools often reject long, detailed responses. P14 continued: \textit{``Our GenAI employs shortened psychological assessment scales tailored for users. We streamlined these scales because users with potential psychological concerns often exhibit impatience during AI interactions. The original scales used in clinical or institutional settings are less intuitive for our audience, and lengthy AI-administered assessments can trigger resistance or anxiety in some users.''}  This disconnect underscores that domain expertise in conventional therapy does not seamlessly transfer to AI therapy design. Psychological scales developed for human-administered assessments may be ineffective—or even harmful—when delivered by AI agents.

This inertia poses even greater risks for vulnerable user groups. To address this, P6 and P7 emphasized the need for stricter AI risk management protocols. For instance, P6, working on a K–12 AI education system, advocated for broadening the definition of harmful content to include subtle or misleading information: \textit{``We usually do not use `harmful' to describe such content. We use ‘not good’ to describe it, but we also aim to red team against such misleading content for our young student users.''} User research can expand the scope of red teaming to better reflect the needs of vulnerable populations. P7 shared how older adults, when exposed to false information through AI-driven weather or disaster alerts, often lose trust in the technology:  \textit{``We conducted a user survey and found that if an elderly person encounters false information through this technology, they are unlikely to trust it again after that initial experience. Moreover, their skepticism or even bias against the technology will likely intensify, potentially reaching a point where the harm is irreparable.''}


\subsection{Red Teaming as Marginalized Labor: Mental Burdens and Structural Inequities}

We found that red teaming generative AI introduces more severe psychological risks than red teaming traditional software systems. Unlike conventional software, generative AI requires continual moderation and feedback on vast volumes of model-generated content—much of it potentially harmful. Because these models rely on human feedback mechanisms (e.g., reinforcement learning from human feedback), red teamers are often tasked with labeling unsettling, inappropriate, or violent outputs. P3 noted that while small datasets can be manually reviewed, the scale of GenAI data has pushed companies to outsource content moderation to large third-party teams tasked with labeling harmful content—\textit{``a massive, third-party team contracted solely for compliance review.''} In this section, we argue that red teaming GenAI is a marginalized form of labor: it is structurally sidelined in organizations, disproportionately burdens its workers psychologically, and is increasingly performed by individuals from marginalized groups.


\subsubsection{Organizational Marginalization of Red Teaming}
Organizational dynamics contribute to red teaming’s marginal status. First, red teams are consistently small in size (see Table 1). Second, their counterparts—the blue teams responsible for defense and mitigation—are usually larger and more resourced. This imbalance in team size translates into a power asymmetry during internal processes. As P7 explained, red team members often feel socially isolated and discouraged:
\begin{quote}
\textit{``During the company's internal product testing, there are usually more blue team members involved, so throughout the process, the red team faces significant (social) pressure. It's something that can't really be avoided. […] At the beginning, our boss implied us do not be that harsh to the blue team. It is kind of a warn to us.''}
\end{quote}

Red teaming efforts are also constrained by unequal access to computational resources across the industry. Dominant firms like Microsoft or Baidu, with early-mover advantages and government relationships, allocate more GPUs and staffing toward security. In contrast, smaller or late-entry firms often deprioritize red teaming to focus on market competitiveness: \textit{``...Microsoft is already in a leading or monopolistic position in the industry, especially since it is an investor of ChatGPT. As a result, they place a strong emphasis on security. Other companies, on the other hand, are in a position of trying to catch up, so their understanding of security is simply about avoiding mistakes and ensuring user traffic. […] Baidu also has the first-mover advantage, as they built earlier relationships with the government. They don’t need to worry about the GPU''} (P3).

\subsubsection{Psychological Toll and Lack of Institutional Support}
Red teamers consistently reported a lack of institutional acknowledgment of the psychological toll associated with evaluating harmful content. Most companies offered no formal psychological support; team members relied on one another for informal care or simply endured silently: \textit{``The company might give us some guidance during meetings, but when it comes to this, we have to take responsibility for maintaining our own mental health''} (P8). In some cases, red teamers felt complicit in one another’s suffering, underscoring a collective emotional burden: \textit{``I feel like I’m also contributing to the mental harm [of my colleagues]. Everyone just tries to stay strong.''}

The mental toll is particularly acute for red teamers who engage with image and video content, especially when tasked with labeling graphic or pornographic materials for model training. As P5 described it:
\begin{quote}
\textit{``Some tasks can indeed be quite detrimental to mental and physical well-being, such as repetitive and extensive labeling work. This kind of work is mentally exhausting, similar to participating in psychological experiments for one or two hours, which, like certain professions, can be harmful to one's health. […] The reason we label harmful content is essentially to provide training data for these so-called automated labeling tools. The question of whether automated labeling tools can protect red team members is somewhat like the `chicken or the egg' dilemma.''}
\end{quote}

\subsubsection{Gendered and Marginalized Labor Dynamics in Red Teaming}
Red teaming is frequently carried out by students, recent graduates, and women—groups with limited organizational authority or recognition. In a government project (P7), six junior red teamers were assigned to test outputs, while seventeen senior engineers constituted the blue team. In another case (P4), the red team was composed entirely of four part-time interns.

Beyond assignment, marginalized workers are subtly steered into harmful content moderation over time. Initially, teams responsible for AI-generated pornography detection were predominantly male. However, male red teamers who expressed discomfort were either reassigned or left the team. Female red teamers, perceived by managers as more ``stable'' and less likely to quit, gradually assumed responsibility for these tasks:
\begin{quote}
\textit{``They (managers) changed the male red teamers to female red teamers […]. At first when I just got into this red team, at that time male red teamers were more than female red teamers. […] But I remember that my (male) senior, who resigned before I came here and introduced me here, said that looking at the porn content generated by AI can indeed impact his life, particularly his sex life. […] Not just him, our manager also said that this kind of work can impact physical health for them (male red teamers). So now, gradually all the red teaming against AI porn content is done by females. If there are female red teamers, they will all be introduced to here.''} (P3)
\end{quote}

This gendered division of labor becomes further normalized through seemingly minor yet telling organizational cues. For instance, when presenting AI-generated pornographic outputs, male presenters elicited discomfort, while female presenters did not: \textit{``[…] when male colleagues presented something related to pornography detection or identification, […] If a male colleague were the one showing it (to red teamers), it would feel slightly uncomfortable […]. However, if a female colleague presented the same material, the atmosphere in the room would be much more normal''} (P3). Even when male red teamers raised concerns about the discomfort and gendered implications of these presentation dynamics, their input was often dismissed or reframed as ``cultural misunderstandings'':\textit{ ``Generative AI technology originates from Western countries […] where, culturally, women are perceived to be more open—more comfortable with sexual content'' }(P16). These patterns reveal how harmful labor assignments—such as presenting and processing pornographic AI outputs—gradually become gendered expectations. \textbf{This normalization is reinforced not only by managerial choices but also by unspoken team dynamics that shape whose discomfort is acknowledged and whose is dismissed.}

\section{Discussion}
Our findings show that red teaming GenAI systems involves multiple internal and external stakeholders, reflecting its growing organizational significance. Yet its implementation is constrained by three organizational dynamics: resistance, inertia, and mediocracy. These dynamics rigidify red teaming workflows, limiting its ability to uncover or address harms. Beyond time and profit pressures, resistance arises from red teaming’s disruptive role—challenging product roadmaps and hierarchies in decision making processes (e.g., who gets to set the roadmap or reject findings). Red teaming in organizational AI safety fits Christensen’s definition of a disruptive technology—challenging existing processes and priorities—but its absence of direct innovation or profits often marginalizes its value ~\cite{christensenInnovatorsDilemmaRevolutionary2011a}. Ultimately, our findings illustrate red teaming as essential yet structurally sidelined labor in organizations.

We propose embedding red teaming throughout the development cycle and integrating user research into red teaming practices. Early involvement can surface harms before sunk costs escalate, allowing for more flexible design changes to GenAI systems or development workflows. We also examine which users—malicious and vulnerable—should be included in user research to better capture real-world risks. Lastly, we suggest that automated red teaming tools may help red teamers legitimize their findings by reframing them aligning organizational values. The following discussion draws on our empirical findings and connects them to existing work in CSCW, responsible AI, and organizational studies to elaborate on these directions.

\subsection{Red Teaming for GenAI as an Organizational Practice}
In our findings, the workflow of red teaming in GenAI development typically involves four interlinked steps: hypothesizing the risks, designing test strategies, conducting adversarial evaluations (i.e., testing and identifying specific harms or vulnerabilities), and proposing mitigation approaches. While these stages may not appear uniformly across all organizations, together they reflect an evolving and systematized red teaming practice. 

When hypothesizing risks, red teaming researchers in academia often work independently, guided by domain expertise or prior literature. In contrast, industry red teamers identify vulnerabilities through informal cues from developers (e.g., P5), direct discussions with clients (e.g., P8), engagement with aggressive users (e.g., P7), or by passively following checklists provided by managers (e.g., P10). These activities show that practical red teaming goes beyond predefined harm categories or attack taxonomies ~\cite{rawat2024attack}, instead relying on contextual understandings of the defensive mechanisms and user behaviors—ranging from malicious actors to vulnerable populations. These context-sensitive practices also extends Suresh et al.’s argument that red teaming has yet to fully engage users in collaborative ways ~\cite{suresh2024participation}, by emphasizing the importance of involving malicious users to better understand real-world threats.

At the same time, our findings reveal limitations in relying on harm and attack taxonomies alone. While such frameworks help structure comparisons across societal impacts ~\cite{rawat2024attack}, they are rarely sufficient in practice—especially in industry contexts, where red teamers are seldom the sole decision-makers. Red teaming plans are often negotiated with developers and clients, particularly in third-party contexts (e.g., P8), where organizational priorities constrain scope. Moreover, these taxonomies often lack the flexibility needed for dynamic adaptation. While they are intended to support comparison and systematic testing, their early review by defense teams can inadvertently constrain red teamers’ strategies. Specifically, when defense teams are aware of taxonomy-based risks in advance, they often preemptively reinforce systems against those patterns—making red teamers’ use of known categories less impactful. As a result, red teamers are frequently encouraged, either explicitly or implicitly, to prioritize novelty over iterative refinement, with an emphasis on creating something `new' rather than thoroughly testing known vulnerabilities. As P13 described, the red team was told to come up with new strategies and avoid reusing existing ones, illustrating how taxonomies can shift from flexible guides to rigid scripts—restricting red teamers’ ability to explore risks systematically. In parallel, the development of test strategies is often siloed within the red team or shaped by select domain experts—such as psychologists, as in P14’s case. Yet these strategies are rarely co-developed with users or broader stakeholders, revealing a disconnect between technical expertise and the lived realities of those affected.

In the context of formal adversarial tests, GenAI systems have significantly shifted the scope of red teaming. Unlike traditional practices that target closed-source and closed-loop systems to gain access or control, GenAI’s open-ended interactivity blurs the lines, bringing user–system data flows—whether from malicious prompts or unintended outputs—into focus. Our findings show that tasks like content moderation, data labeling, and harm-based model training often require large-scale collaboration, including outsourced labor (e.g., P3). Furthermore, our findings indicate that the work of red teamers and that of blue teamers is not clearly separated.  Red teamers are often expected to go beyond identifying risks—they must also propose solutions or mitigation strategies for blue teams or engineers responsible for implementing defenses (e.g., P9, P11). Together, these patterns suggest that red teaming in GenAI is not just a technical task, but an embedded organizational practice. It demands rethinking traditional team structures, expanding who participates in red teaming work, and clarifying the institutional responsibilities for managing risk across the entire development pipeline.

\subsection{Three Organizational Dynamics Causing Rigidity in Red Teaming}
Recent research starts to see red teaming the GenAI systems as not just a technical task, but a form of labor embedded within social, institutional, and political contexts ~\cite{gillespieAIRedteamingSociotechnical2025, singhRedTeamingPublicInterest2025}. In the previous research about red teaming, the team dynamics between red team and blue team has been a central focus related with this labor perspective. By design, red and blue teams are positioned in adversarial roles, where the red team simulates attacks, and the blue team defends against them. On the team level, their goals are totally opposite. But from the organizational level, the red team and blue team all aim at enhancing the safety of the computational system. However, as red teamers \textit{``generally grade the defenders on their ability to respond to attacks, the outcome often causes blame shifting, distrust, and lowered morale among team members''} ~\cite{pottiEmotionalTollRed2023}. Uncertainty in the planning phase can introduce team dynamics ~\cite{kraemerRedTeamPerformance2004}. To solve this, Abbass et al. ~\cite{abbass2011computational} suggests having the red team reproduce the malicious users' motivations, intentions, behaviors and anticipated actions, to encourage blue team to \textit{``appreciate the dynamics of how Blue and Red interact and gain an understanding of the space in which the dynamics may unfold and evolve.''}

Our paper's investigations expand the extant observations of red/blue team dynamics to within-organization level, and identified three organizational dynamics when red teaming is implemented for GenAI systems: organizational resistance, organizational inertia, and organizational mediocracy. 

\textit{\textbf{Organizational Resistance}} When red teaming uncovers critical vulnerabilities that require substantial changes to a product, organizational resistance may arise as the organization refuses to adopt the necessary change\footnote{Here, resistance refers to ``any conduct that serves to maintain the status quo in the face of pressure to alter the status quo'' ~\cite{zaltmanStrategiesPlannedChange1977}.}. In the red teaming context, the organizational resistance is not abstract but enacted through key organizational actors who, either implicitly or explicitly, oppose or delay implementation of the recommended changes. 

These organizational dynamics echo prior findings in CSCW. Drawing on this body of work, we analyze how such resistance unfolds in GenAI development environments. Similarly, in AI ethics discussions, time pressure can lead to discursive closure, and the push for solutionism narrows ethical critiques to only those issues that are seen as addressable~\cite{widderPowerPlayInvestigating2024} or upon short-term goals~\cite{Rakova2021WhereRAI, Wong2021Tactics}. Beyond time constraints and a focus on profit margins, red teaming results are often resisted by organizations due to two unique factors: the substantial sunk costs incurred during early development stages, and the perception of red teams as a disruptive force. As P7 described, the pressure red team faces in adversarial tests extends beyond the dynamics between blue team and red team. The red teaming results were often viewed as \textit{``trouble later on.''}
 
Similarly, when the architectural flaw were surfaced, P8 said how clients responded with fury. Rather than simply identifying vulnerabilities, effective red teaming challenges the existing organizational order—including architectural decisions, development priorities, and internal lines of accountability.  In this sense, red teaming echoes with Clayton Christensen’s concept of \textit{disruptive technology} ~\cite{christensenInnovatorsDilemmaRevolutionary2011a}. Drawing on the example of the cable excavator’s decline and the hydraulic excavator’s rise, Christensen argues that large companies often struggle to adopt disruptive technologies because doing so may undermine their existing business models ~\cite{christensenInnovatorsDilemmaRevolutionary2011a}. Because disruptive technologies rarely target large markets at the outset, they often address emerging or unrecognized needs. This uncertainty makes it difficult for companies to identify or respond to these unfulfilled demands. Jeroen Hopster argues that in addition to disrupting the business model, disruptive technologies also disrupt social relations and institutions of organization ~\cite{hopsterWhatAreSocially2021}. From this aspect, red teaming is a disruptive force deliberately examining the underlying assumptions and challenges the business model, social relations and internal institutions (e.g. P8 faced strong criticism and was told to pause red teaming after challenging the architecture of GenAI system). In other words, the disruptive nature of red teaming carries emotional costs, as red teamers must navigate institutional pushback and strained social dynamics.

As the contrarian team, our findings illustrate that red team constitutes a form of essential yet marginalized labor within organizations—often carried out by individuals from already marginalized groups. This aligns with prior research on responsible AI practices (RAI), which finds that marginalized individuals tend to shoulder disproportionate amounts of invisible and emotional labor when advocating for fairness and accountability ~\cite{dengInvestigatingPracticesOpportunities2023, deng2023understanding, widderPowerPlayInvestigating2024}. These individuals, particularly women, junior employees, and those in non-engineering roles, are often the first to notice sensitive risks such as racial or gender bias ~\cite{deng2023understanding}, yet they frequently encounter organizational resistance or social discouragement when voicing such concerns ~\cite{widderPowerPlayInvestigating2024}. They report bearing disproportionate emotional labor in advocating for fairness—often without meaningful institutional support ~\cite{dengInvestigatingPracticesOpportunities2023}.

We observe similar dynamics in red teaming practices. Female red teamers, for instance, demonstrated heightened awareness of fairness-related harms (e.g., P3), whereas male peers sometimes dismissed such concerns (e.g., P16). These patterns parallel findings from studies of content moderation on social media platforms, where marginalized workers disproportionately encounter and absorb the harms of toxic content to improve user experiences for others ~\cite{stackpoleContentModerationTerrible2022, arsht2018human}.
Our findings further show that junior red teamers (e.g., P8) often face greater resistance when attempting to have their findings taken up, adding emotional strain to their work. In GenAI red teaming, the harms encountered are not abstract but often vividly rendered—for example, in image- or video-based outputs—which exacerbates the emotional toll. This mirrors the experiences of content moderators in under-resourced countries, many of whom report PTSD after prolonged exposure to violent or discriminatory content ~\cite{perrigo150AIWorkers2023, dachwitzDataWorkersInquiry2024}. As P5 noted, red teaming can feel like ``sitting through a two-hour-long psychological experiment,'' underscoring the urgency of supporting red teamers, particularly when they belong to the very groups most affected by the harms they audit (e.g. women auditing the content of bias on women).

\textit{\textbf{Organizational Inertia}} Organizational Inertia occurs when red teaming fails to play its contrarian role—finding few or no risks—not because the system is truly safe, but due to the organizational settings that discourage rigorous scrutiny. In such cases, red teamers may be distracted from uncovering real harms in GenAI systems. This reflects a broader organizational stagnation, where companies remain overly reliant on stable products, production methods, and policies, even when change is warranted~\cite{huang2013overcoming}. Observing the response of newspaper organizations to the rise of digital media, Gilbert found organizational inertia can appear in the form of rountine inflexibility, which refers no change in organizational processes and the procedures of using the invested resources ~\cite{gilbertUnbundlingStructureInertia2005}. Our findings show two forms of organizational inertia in red teaming. In the case of P5, both the red teamer and the blue teamer focused exclusively on identifying vulnerabilities within the software development process. Thus red team can easily fall into the defense trap set by blue team (e.g. Honeypot, by which attackers are hampered to attack those objectives set by defense on purpose). However, in the case of P13, red teaming exhibited exhibited a form of organizational inertia, characterized by a self-limiting focus—red teamers avoided exploring strategies that blue teams had already accounted for. Previous CSCW research found organizational dynamics can create an informal license effect to discourage people's perceived legitimacy to critique on ethical issues ~\cite{widderPowerPlayInvestigating2024}. Our findings show the license do not limit to who are able to critique, also apply to who can decide using what strategy to attack. To counter this inertia, one way is involving publics outside the organization, who can introduce fresh perspectives and novel threat models. Another approach is to leverage risk taxonomies, which can help red teamers compare, combine, and iterate on different strategies to identify blind spots in the system's defense ~\cite{rawat2024attack}.

\textit{\textbf{Organizational Mediocracy}} Organizational mediocracy describes the situation when harms in GenAI system are successfully identified and the organization is willing to address the harms, however, the organization only partially eliminate the worst cases, rather than aiming directly at safety. This can occur when red teaming, which is meant to promote GenAI safety aligned with regulators’ intentions, is reduced to a compliance-driven effort aimed at avoiding punishment or escaping regulatory attention. One interviewee illustrated this attitude with irony: \textit{``We won’t be the least compliant in AI safety, but we’ll aim to be the second least''} (P3). To counter organizational mediocracy, we emphasize the importance of leadership buy-in and clear principles for balancing ethical commitments with business imperatives. The mediocracy is also shaped by competitive dynamics in the GenAI industry, particularly for companies striving to catch up with market leaders. As Ali et al.~\cite{Ali2023Walking} argue, resource constraints on AI ethics efforts often stem from a lack of institutional support for ethics practitioners. Our findings expand on this by pointing to structural imbalances in computational resource allocation—where first movers like big tech companies (e.g., Microsoft) enjoy privileged access~\cite{cheon2023powerful} to GPUs, while newer entrants struggle to secure sufficient capacity for both innovation and safety testing. In a highly competitive GenAI industry, companies struggling to catch up often prioritize allocating their limited computational resources (e.g., GPUs) to model capability over safety. As a result, they adopt a ``bare-minimum'' mindset reflective of a mediocracy—treating red teaming less as a safety-critical task and more as a symbolic gesture to satisfy compliance demands.


\subsection{The Potential of User Research \textit{for} Red Teaming}
The above suggests that adversarial testing remains in its early stages and lacks a robust framework for enabling, as Wang et al. put it, \textit{``UX practitioners to either lead such efforts themselves or incorporate their results into UX design and evaluation''} ~\cite{wangDesigningResponsibleAI2023}. More specifically, unlike other responsible AI practices, red teaming as an adversarial lens is not yet integrated into the early design and development stages ~\cite{wangDesigningResponsibleAI2023}. Our findings align with Suresh et al.’s work, suggesting that red teaming has yet to establish meaningful collaboration with users ~\cite{suresh2024participation}. In our study, the lack of direct engagement led red teamers to develop only a vague understanding of who their vulnerable users were—for example, referring broadly to \textit{``teenagers whose psychological status can be impacted''} (P17) or \textit{``only consumers [of our e-commerce platform]''} (P11). In RAI practices, aligning fairness concerns with business interests can help gain leadership support ~\cite{Wong2021Tactics}. Red teamers, however, often lack direct connection to users, which limits their ability to advocate for user-centered harms. We therefore discuss potential ways to involve red teamers in user research.

To better connect red teamers with users, one potential way is to integrate red teaming throughout the entire software development lifecycle. As Lucaj et al.'s findings suggests, red team methods should be combined with audit-trail and be applied from the early-design phases to the post-market monitoring ~\cite{lucaj2023ai}. Similarly, Yildirim et al.’s work shows that lacking a human-centered perspective during early problem formulation can constrain the design space. Once a dataset is collected and a machine learning model is built, reframing the model to solve the ethics problem becomes significantly limited and costly ~\cite{Yildirim2023Investigating}. Our findings unpack how rigid design decisions—such as fixed model architectures or narrowly defined use cases—made during early development can hinder further iteration, even when red teaming reveals clear ethical or functional vulnerabilities in the system. In addition to making iteration costly, sunk costs also contribute to organizational tensions, as red teamers may be seen as challenging established decisions that are costly to reverse. Building on Yildirim et al.’s work, future research should explore how red teaming methods can utilize guidelines to educate and communicate RAI principles with their organizations and clients ~\cite{Yildirim2023Investigating}. In addition, \textbf{integrating red teaming earlier in the design and development cycle may reduce perceptions of red teamers as disruptive actors, particularly because the system is still in flux and more receptive to critique—thereby easing emotional labor demands.
}

Another key question is which users red teamers should involve in their processes. Singh et al. advocate red teaming should consider and involve the very relevant publics and proactively view the end users as the last line of defense ~\cite{singhRedTeamingPublicInterest2025}. By involving publics, red team will gain more diversity, draw on people's personal experiences to handle the subjective nature of what is harmful, and connects GenAI systems more closely to real-world contexts ~\cite{singhRedTeamingPublicInterest2025}. Involving publics also provides red teams with opportunities to build external alliances, helping to compensate for the limitations of red teaming being confined to pre-deployment stages—where it lacks the motivational triggers for responsible AI responses that might otherwise come from media exposure or public scandals. External groups are often more free to speak up than internal ones, which can indirectly amplify the voice and power of red team members ~\cite{Rakova2021WhereRAI}. Beyond vulnerable users, the malicious users should also be incorporated into user research, as they represent potential threat actors who may exploit GenAI systems to harm others (e.g., P7's team invited adversarial users to participate in testing). This perspective has been largely overlooked in prior responsible AI (RAI) discussions, which often assume users are well-intentioned. Future research should further examine the role of malicious users in both RAI practices and red teaming.

UX professionals may adopt soft resistance by redefining ethics concerns into business concerns through user research ~\cite{Wong2021Tactics}. We found this tactic is difficult for red teamers to adopt due to their disconnection from users. For instance, P10’s role was limited to a small, compartmentalized portion of a pre-defined keyword list—just one-fourth of the total keyword set—leaving little room for broader ethical reflection. Despite this constraint, we see potential in automated red teaming tools as a middle ground between red teams and organizations. These tools offer partial, rather than full, automation—allowing red teamers to retain agency in defining harms, adjusting inputs, and designing strategies. As P12 suggests, human red teamers will continue to guide the cold-start phase, which may present opportunities to reintroduce soft resistance by shaping the direction and framing of early red teaming efforts. However, incorporating user research adds to the significant cognitive and emotional burden red teamers already face. This raises an important question: what should the long-term trajectory look like for red teamers?

\section{Conclusion}
In response to the growing risks posed by generative AI, this study examined how red teamer practice red teaming across different organizational contexts. We show that red teaming for generative AI is deeply shaped by organizational dynamics—namely, resistance, inertia, and mediocracy. These forces limit not only what harms are identified, but also who gets to identify them, often sidelining both red teamers and user perspectives. To make red teaming more effective and more inclusive, we call for its integration across the GenAI development lifecycle and recognizing red teamers as sociotechnical actors embedded within organizational structures. Our findings provide a grounded understanding of red teaming as collaborative, labor-intensive, and politically situated work, underscoring its role not as an isolated security task, but as a core organizational practice in building responsible AI systems.

\bibliographystyle{ACM-Reference-Format}
\bibliography{reference}

\end{document}